\documentclass[reprint,amsmath,amssymb,aps,prb,twocolumn,float]{revtex4-2}
\usepackage{graphicx}
\usepackage{graphics}
\usepackage{dcolumn}
\usepackage{bm}
\usepackage{amsfonts}
\usepackage{amssymb}
\usepackage{float}
\usepackage{xcolor}
\usepackage{soul}
\usepackage{natbib}
\usepackage[normalem]{ulem}
\usepackage{soul}

\usepackage{caption}
\usepackage{subcaption}

\begin{document}

\title{\textit{Ab initio} study on magnetism suppression, anharmonicity, rattling mode and superconductivity in Sc$_6M$Te$_2$ ($M$=Fe, Co, Ni)}

\newcommand{\RCAST}{Research Center for Advanced Science and Technology, University of Tokyo, 4-6-1 Meguro-ku, Tokyo, 153-8904, Japan}
\newcommand{\Riken}{RIKEN Center for Emergent Matter Science, 2-1 Hirosawa, Wako 351-0198, Japan}
\newcommand{\QPEC}{Quantum Phase Electronics Center, University of Tokyo, Tokyo 113-8656, Japan}
\newcommand{\AP}{Department of Applied Physics, University of Tokyo, Tokyo 113-8656, Japan}
\newcommand{\NIMS}{Center for Basic Research on Materials, National Institute for Materials Science, Tsukuba, Ibaraki 305-0047, Japan}
\newcommand{\MaDIS}{Research and Services Division of Materials Data and Integrated System, National Institute for Materials Science, Tsukuba, Ibaraki 305-0047, Japan}
\newcommand{\ESISM}{Center for Elements Strategy Initiative for Structural Materials, Kyoto University, Sakyo, Kyoto 606-8501, Japan}
\newcommand{\NTU}{Department of Physics and Center for Theoretical Physics, National Taiwan University, Taipei 10617, Taiwan\looseness=-1}
\newcommand{\NCTS}{Physics Division, National Center for Theoretical Sciences, Taipei 10617, Taiwan\looseness=-1}

\author{Ming-Chun Jiang$^{1,2}$}\email{ming-chun.jiang@riken.jp}
\author{Ryota Masuki$^{3}$}
\author{Guang-Yu Guo$^{1,4}$}          
\author{Ryotaro Arita$^{2,5}$}        

\affiliation{$^1$\NTU}
\affiliation{$^2$\Riken}
\affiliation{$^3$\AP}
\affiliation{$^4$\NCTS}
\affiliation{$^5$\RCAST} 

\date{\today}

\begin{abstract}
We perform a systematic \textit{ab initio} study on phonon-mediated superconductivity in the transition-metal-based superconductors Sc$_6M$Te$_2$ ($M$ = Fe, Co, Ni).
Firstly, our charge analysis reveals significant electron transfer from Sc to $M$ due to the substantial difference in the electronegativity, filling the 3$d$ orbitals of $M$ and suppressing magnetic instability.
Secondly, we show that Sc$_6$FeTe$_2$ exhibits strong lattice anharmonicity.
Moreover, for $M =$ Fe and Co, we find low-frequency soft phonon bands of $M$ which can be interpreted as ``rattling phonons" in the framework formed by Sc. While not observed in the case of $M=$ Ni, the rattling phonons give rise to a prominent peak or plateau in the Eliashberg spectral function and enhance the pairing instability.
By reproducing the experimental trend of superconducting transition temperatures, our study underscores the potential of designing phonon-mediated superconductors by strategically combining non-superconducting and magnetic transition-metal elements.
\end{abstract}

\maketitle

\section{INTRODUCTION} 

Identifying common characteristics among known superconductors bears significance for  designing novel superconductors~\cite{Fisk2009}.
Despite the abundance of identified superconductors, platforms remain scarce, where superconductivity manifests across a series of materials. 
This is particularly true in compounds involving $d$ electrons~\cite{Bednorz1986,Kamihara2008,Hsu2008,Heintz1989,Maeno1994,Takada2003,Yonezawa2004,Ortiz2020,Li2019,Sen2020,Shinoda2023}.
Recently, seven materials from a family of $d$-element-rich compounds, Sc$_6M$Te$_2$, have been identified as possessing bulk superconductivity~\cite{Shinoda2023}.
These materials have the Zr$_6$CoAl$_2$-type structure, with $M$ encompassing 3$d$, 4$d$, and 5$d$ elements.
Notably, the compounds with $M=$ Fe, Co, and Ni show a significant material dependence in their superconducting transition temperatures ($T_c$'s) of 4.7 K, 3.6 K, and 2.7 K, respectively. 
Interestingly, the pairing mechanism is yet to be fully understood:
In particular, it is unclear why superconductivity occurs in compounds containing magnetic elements and why $T_c$ is
highest with $M=$ Fe and lowest with $M=$ Ni.

To address these questions, we perform a systematic \textit{ab initio} study on the electronic structures, phonon properties, and the electron-phonon coupling (EPC) for Sc$_6M$Te$_2$ ($M$ = Fe, Co, Ni). We first perform the charge analysis based on the electronic structure, which shows an electron transfer from Sc to $M$. As a result, the $3d$ orbitals of $M$ are almost fully occupied, which suppresses the local magnetic moments on $M$. By reproducing the chemical trend of $T_c$, we conclude that the target materials are phonon-mediated superconductors. 
In addition, we show that the higher $T_c$ in Sc$_6$FeTe$_2$ and Sc$_6$CoTe$_2$ can be ascribed to the strong EPC from the low-frequency phonon modes dominated by the displacements of $M$. 
 
The crystal structure of Sc$_6M$Te$_2$ is characterized by the Sc cages encapsulating the isolated $M$ atoms, indicating the system to be host(Sc)-guest($M$) compounds. Due to such structural arrangement and the emergence of low-frequency phonons with single-atom vibrations, we would interpret such phonons as ``rattling phonons". This phenomenon is often observed in materials where large freedom of motion is given to guest atoms within periodic cages with examples of skutterudites~\cite{Nolas1996,Sales1997}, pyrochlores~\cite{Hiroi2005,Yoshida2007,Dahm2007}, clathrates~\cite{Nolas1996}, or brownmillerites~\cite{Rykov2004}.

Non-parabolic behavior of the potential energy surface due to atomic displacements, namely the lattice anharmonicity, can occur for a rattling vibrational behavior~\cite{Wei2021}. While the effect of anharmonicity on superconductivity has been extensively studied~\cite{Yu1984,Errea2013,Errea2014,Errea2015,Errea2016}, 
the combining effect of rattling modes and anharmonicity on superconductivity was discussed for the $\beta$-pyrochlore oxides $A$Os$_2$O$_6$~\cite{Nagao2009} and intermetallic cage compounds $M$V$_2$Al$_{20}$~\cite{Winiarski2016}. Since rattling modes provide significant atomic displacement at the flat potential energy surface, the attractive interaction for the Cooper pair can be large and thus enhance $T_c$~\cite{Nagao2009,Winiarski2016}. 
Although there are some cases where the rattling modes have only a moderate impact on superconductivity~\cite{Uzunok2019}, we find that taking advantage of the strong EPC of the soft rattling phonons is an effective strategy to enhance superconductivity in Sc$_6M$Te$_2$.

The rest of this paper is organized as follows. 
Secion II describes the theoretical methods, the crystalline structure of Sc$_6M$Te$_2$, and the computational details. 
Section III presents the calculated electronic structures, including the orbital projected density of states (DOS), band structures, and the Fermi surface. 
Then, we show the phonon band structures, atom-projected phonon DOS, and the corresponding Eliashberg spectral function to quantify the EPC strength. 
Further discussions on the suppression of magnetism and the emerging superconductivity in Sc$_6M$Te$_2$ are provided in Sec. IV. 
Finally, the conclusions drawn from this work are summarized in Sec. V.

\section{Theory and computational methods}

\subsection{Electron phonon coupling}

Based on density functional theory (DFT), we can use density functional perturbation theory (DFPT) to calculate the EPC matrix element~\cite{Giustino2017,Wierzbowska2006,Baroni2001}.
\begin{align}
\label{eq:g}
g_{mn,\nu}(\textbf{k,q})&=\sum_{\kappa\alpha}\sqrt{\frac{\hbar}{2\omega_{\nu\textbf{q}}}}\frac{e_{\kappa\alpha\nu\textbf{q}}}{\sqrt{M_\kappa}} \nonumber\\ 
&\times\left\langle\psi_{m\textbf{k}+\textbf{q}}\left\vert \sum_{p}\frac{\partial V}{\partial u_{\kappa\alpha\textbf{R}_p}}\frac{e^{i\textbf{q} \cdot \textbf{R}_p}}{\sqrt{N_p}}\right\vert\psi_{n\textbf{k}}\right\rangle,
\end{align}
where $\omega_{\nu\textbf{q}}$ is the phonon frequency of mode $\nu$ and crystal momentum $\textbf{q}$; $e_{\kappa\alpha\nu\textbf{q}}$ is the polarization vector of site $\kappa$ in the Cartesian direction $\alpha$; $M_{\kappa}$ is the mass of the atom at site $\kappa$. 
$\psi_{n\textbf{k}}$ is the Kohn-Sham orbital with the band index $n$ and wave vector \textbf{k}. $V$ is the self-consistent potential, and its first-order derivative with respect to $u_{\kappa\alpha\textbf{R}_p}$ is calculated using DFPT~\cite{Baroni2001, Wierzbowska2006,Giustino2017}. Note that $u_{\kappa\alpha\textbf{R}_p}$ is the atomic displacement in the lattice position $\textbf{R}_p$ of the $p$-th unit cell. $N_p$ is the number of primitive cells in the Born-von K\'arm\'an supercell. 

The EPC strength of a specific phonon mode $\nu$ and phonon momentum \textbf{q} can then be obtained via the Brillouin zone (BZ) integration~\cite{Giustino2017,Wierzbowska2006,Allen1972}:
\begin{align}
\label{eq:lambda}
\lambda_{\nu\textbf{q}}=\frac{1}{N(\epsilon_\text{F})\omega_{\nu\textbf{q}}}\sum_{nm}&\int_{\rm BZ}\frac{d\textbf{k}}{\Omega_{\rm BZ}}\left\vert g_{mn,\nu}(\textbf{k,q}) \right\vert^2 \nonumber\\
&\times\delta(\epsilon_{n\textbf{k}}-\epsilon_\text{F})\delta(\epsilon_{m\textbf{k+q}}-\epsilon_\text{F}),
\end{align}
where $N(\epsilon_\text{F})$ is the electronic DOS at the Fermi level $\epsilon_\text{F}$ and $\epsilon_{n\textbf{k}}$ is the state energy at momentum \textbf{k}. $\Omega_{\rm BZ}$ is the volume of the BZ.
We then introduce the isotropic Eliashberg spectral function~\cite{Giustino2017,Wierzbowska2006,Allen1972}
\begin{equation}
\label{eq:a2F}
\alpha^2F(\omega)=\frac{1}{2}\sum_{\nu}\int_{\rm BZ}\frac{d\textbf{q}}{\Omega_{\rm BZ}} \omega_{\nu\textbf{q}}\lambda_{\nu\textbf{q}}\delta(\omega-\omega_{\nu\textbf{q}}),
\end{equation}
from which the EPC constant $\lambda$ is obtained as 
\begin{equation}
\label{eq:lama2F}
\lambda = 2 \int \frac{\alpha^2 F(\omega)}{\omega} d\omega. 
\end{equation}

Through the celebrated McMillan-Allen-Dynes (MAD) formula~\cite{McMillan1968,Dynes1972,Allen1975}, we can evaluate the superconducting temperature $T_c$ by
\begin{equation}
\label{eq:MAD}
    T_c = \frac{\omega_{\log}}{1.2} \exp\left[-\frac{1.04(1+\lambda)}{\lambda-\mu^*(1+0.62\lambda)}\right],
\end{equation}
in which the prefactor includes the logarithmic average phonon frequency of the spectrum $\omega_{\log}$ where
\begin{equation}
\label{eq:logW}
    \omega_{\log}=\exp\left[\frac{2}{\lambda}\int\frac{d\omega}{\omega}\alpha^2 F(\omega)\log{(\omega)}\right]
\end{equation}
and $\mu^*$ is the pseudo-Coulomb parameter introduced to account for the screened electron-electron interaction. 

\subsection{Self-consistent phonon theory}
\label{subsec:theory_SCPH}
The self-consistent phonon (SCPH) theory~\cite{Hooton1958, Tadano2015, Tadano2018} is a mean-field theory of the lattice anharmonicity, in which we solve a self-consistent equation in terms of the renormalized effective frequencies $\Omega_{\nu \textbf{q}}$.
\begin{equation}
\label{eq:SCPH}
\Omega^{2}_{\nu \textbf{q}}=\omega_{\nu\textbf{q}}^2
+\frac{1}{2}\sum_{\nu_1 \textbf{q}_1}F_{\textbf{q}\textbf{q}_1,\nu \nu_1}\frac{\hbar[1+2n_B(\Omega_{\nu_1 \textbf{q}_1})]}{2\Omega_{\nu_1 \textbf{q}_1}},
\end{equation}
where $n_B(\omega)$ is the Bose-Einstein distribution function.
$F_{\textbf{q}\textbf{q}_1,\nu \nu_1}=\Phi(\textbf{q}\nu;-\textbf{q}\nu;\textbf{q}_1 \nu_1;-\textbf{q}_1 \nu_1)$ is the fourth-order interatomic force constant (IFC) which is defined in reciprocal representation as~\cite{Tadano2015}:
\begin{multline}
\noindent\Phi(\textbf{q}_1\nu_1;\cdots;\textbf{q}_4\nu_4)=\\
N_p^{-1}\sum_{\kappa\alpha}(M_{\kappa_1}\cdots M_{\kappa_4})^{-\frac{1}{2}} e_{\kappa_1\alpha_1\nu\textbf{q}_1}\cdots e_{\kappa_4\alpha_4\nu\textbf{q}_4}\\
\times\sum_{p_2,\cdots,p_4}\Phi_{\alpha_1\cdots\alpha_4}(0\kappa_1;\cdots;p_4\kappa_4)e^{i(\textbf{q}_2\cdot\textbf{R}_{p_2}+\cdots+\textbf{q}_4\cdot\textbf{R}_{p_4})}.
\end{multline}
where  $\Phi_{\alpha_1\cdots\alpha_4}(p_1\kappa_1;\cdots;p_4\kappa_4)$ is the fourth-order derivative, or the quartic term of the Taylor expansion coefficient, of the Born-Oppenheimer potential energy surface $U$ with respect to displacement $u_{\kappa\alpha\textbf{R}_p}$: 
\begin{equation}
    \Phi_{\alpha_1\cdots\alpha_4}(p_1\kappa_1;\cdots;p_4\kappa_4)=\left.\frac{\partial^4U}{\partial u_{\kappa_1\alpha_1\textbf{R}_1}\cdots\partial u_{\kappa_4\alpha_4\textbf{R}_4}}\right\vert_{u=0}.
\end{equation}
Equation (\ref{eq:SCPH}) can be derived using the Dyson equation with a frequency-independent self-energy~\cite{Tadano2015} or using the variational principle of the free energy~\cite{Tadano2018}. Note that the polarization vectors are assumed not to change by the anharmonic effect in this work. 

As we discuss later in Section~\ref{subsection:Phonon_properties}, we obtain imaginary frequency ($\omega_{\nu\textbf{q}}^2<0$) phonons in the harmonic phonon calculation for Sc$_6$FeTe$_2$.
This is inconsistent with the experiment because the imaginary modes indicate the instability of the identified crystal structure. 
Thus, we use the SCPH theory to obtain a stable phonon dispersion with real frequencies. 
Also, equations (\ref{eq:g}) and (\ref{eq:a2F}) indicate that imaginary frequency phonons would give an ill-defined EPC matrix element and Eliashberg spectral function, which should be handled carefully for phonon-mediated superconductors~\cite{Errea2013,Errea2014,Errea2015,Errea2016,Sano2016,Errea2020}. 
While one may artificially set the imaginary phonon frequencies to zero or their absolute values or introduce a sizeable electronic smearing parameter~\cite{Ramesh2019,Yang2024}, in this work, we employ a treatment that uses the SCPH theory. 
Namely, we obtain the SCPH frequencies $\Omega_{\nu \textbf{q}}$ and replace the harmonic frequencies $\omega_{\nu \textbf{q}}$~\cite{Zhou2018,Chen2023}.
Note that using the SCPH theory is not entirely valid because it considers the finite-temperature renormalization of the phonon frequencies that originate from the anharmonic effect. However, we expect that this treatment is helpful because the SCPH frequencies reflect the overall curvature of the low-energy region of the potential energy surface, while the harmonic calculation is determined only by the curvature at the reference structure. 
Note that we carefully choose the temperature to estimate $n_B(\Omega_{\mu \textbf{q}}$) in the SCPH theory ($T_{\rm SCPH}$) and check that the $T_{\rm SCPH}$-dependence of the calculated $T_c$ is relatively small, which is further discussed in Appendix B.

\subsection{CRYSTAL STRUCTURE AND COMPUTATIONAL DETAILS}

\begin{figure}[tbph] \centering
\includegraphics[width=\linewidth]{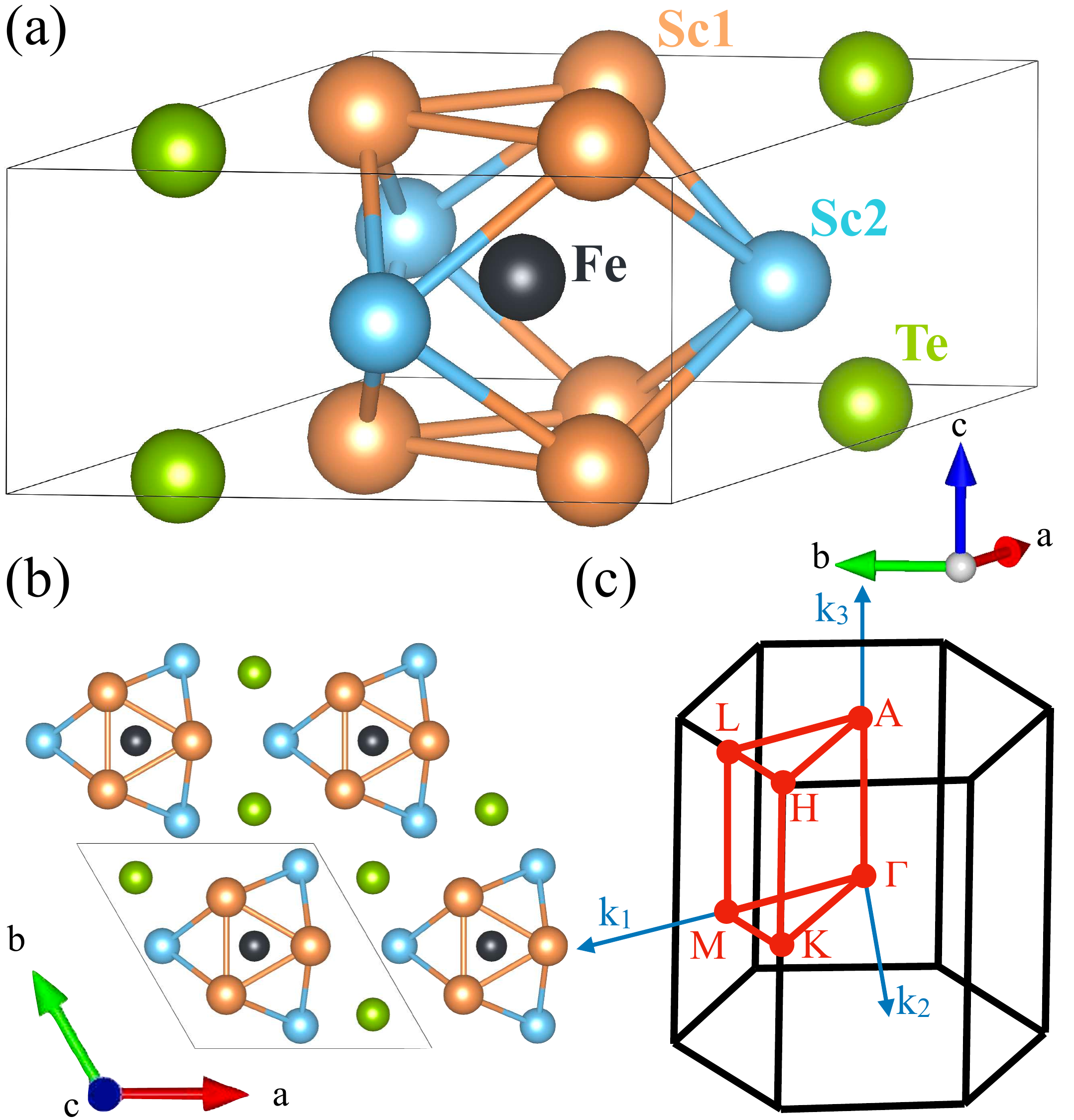}
\caption{Hexagonal crystal structure with space group $P\bar{6}2m$ (No. 189) and Brillouin zone of Sc$_6M$Te$_2$ ($M$= Fe, Co, Ni). (a) Side view and (b) top view of the crystal structure where one unit cell contains one formula unit. (c) The hexagonal Brillouin zone. }
\label{fig:struc}
\end{figure}

The crystal structure of Sc$_6M$Te$_2$ ($M$= Fe, Co, Ni) is hexagonal with the $P\bar{6}2m$ space group (No. 189) which contains one formula unit (f.u.) per unit cell. 
The Sc$_6M$Te$_2$ compounds are an ordered variant of the Fe$_2$P structure, belonging to a large family characterized by the Zr$_6$CoAl$_2$-type structure~\cite{Maggard2000, Chen2003}.
The transition metal elements $M$ are surrounded by the Sc atoms forming tricapped trigonal prisms~\cite{Maggard2000}. 
Such a local Sc cage with $M$ bounded inside stack up along the $c$-direction, which forms a linear chain of $M$. 
The structures are shown in Fig.~\ref{fig:struc}. 
In the current study, we perform full structural relaxation with the results summarized in Table~\ref{table:struc}. 

\begin{table}[htbp]
\caption{Theoretical lattice constants ($a$, $c$, $c$/$a$) of Sc$_6M$Te$_2$ compared with the experimental data~\cite{Maggard2000}. The number in the parenthesis denotes the percentage deviation of the theoretical lattice constants from the experiments.}
\begin{ruledtabular}
\begin{tabular}{c c c c}
Sc$_6M$Te$_2$  &  $a$ (\AA)   & $c$ (\AA)  &  $c$/$a$   \\
\hline
Fe& 7.703 (+0.306\%)& 3.788 (-1.272\%)& 0.492\\
 & 7.6795$^a$& 3.8368$^a$& 0.4996\\
Co& 7.732 (+0.446\%)& 3.743 (-1.123\%)& 0.484\\
 & 7.6977$^a$& 3.7855$^a$& 0.4918\\
Ni& 7.768 (+0.576\%)& 3.729 (-0.972\%)& 0.480\\
 & 7.7235$^a$& 3.7656$^a$& 0.4876
\end{tabular}
\end{ruledtabular}
{$^a$Experimental data taken from Ref.~\cite{Maggard2000}}
\label{table:struc}
\end{table}

The structural relaxation and the calculations of electronic and phonon properties are conducted using DFT, and the corresponding electron-phonon properties are calculated using DFPT~\cite{Baroni2001}, as implemented in the QUANTUM ESPRESSO (QE) code~\cite{Giannozzi2009}. 
All the calculations are performed using the scalar-relativistic projector-augmented wave (PAW)~\cite{Blochl1994} pseudopotential with the generalized gradient approximation (GGA) in the form of Perdew-Burke-Ernzerhof (PBE)~\cite{Perdew1996} from PSLibrary~\cite{Corso2014}. 
A plane-wave cutoff value of 952 eV and a $\Gamma$-centered 12 × 12 × 24 Monkhorst-Pack~\cite{Monkhorst1976} $k$-mesh was used to describe the electronic structure. 
For the DOS calculation, we apply the tetrahedron method~\cite{Blochl1994_tetra}. 
The valence orbital set is $3s^23p^64s^23d^{1,7,8,9}$ for Sc, Fe, Co, Ni, and $5s^25p^4$ for Te. 
The phonon calculations are performed with a $q$-grid of 2 $\times$ 2 $\times$ 2. 
Also, the crystal structure is visualized by VESTA~\cite{Momma2011}, and the Fermi surface is plotted by FermiSurfer~\cite{Kawamura2019}.  

We perform the SCPH calculation for Sc$_6$FeTe$_2$. 
The IFCs are generated using Vienna \textit{ab initio} simulation package (VASP)~\cite{Kresse1993,Kresse1996} and ALAMODE package~\cite{Tadano2015,Tadano2018, Tadano2014}, where we use a 2 $\times$ 2 $\times$ 2 supercell containing 72 atoms. 
The VASP calculations for the force data are performed using DFT with the PAW method~\cite{Blochl1994} and the GGA in the form of PBE~\cite{Perdew1996,Kresse1999}. 
A plane wave cut-off energy of 520 eV is used. For the BZ integration, a $\Gamma$-centered $k$-mesh of 4 $\times$ 4 $\times$ 4 is used for the supercell.       
The harmonic IFCs are calculated by the frozen phonon method with atomic displacements of 0.01 \AA. 
The anharmonic IFCs are computed using the compressive sensing method, which enables efficient calculation of IFCs from a small number of supercell calculations~\cite{Zhou2019}. 
To generate the displacement-force data from which we extract the IFCs, we generate configurations with random atomic displacements using \textit{ab initio} molecular dynamics (AIMD) simulation. 
To sample the low-energy region of the potential surface, we first perform the AIMD in VASP at 300K for 5000 steps with a time step of 1 fs. 
We discard the first 1000 steps due to their dependence on the initial setting of the AIMD simulation and extract 80 snapshots from the rest of the trajectory. 
We further displace the atoms randomly by 0.04 \AA\ to eliminate the correlation between the snapshots. 
DFT calculations of VASP compute the atomic force for each supercell, and up to the fourth-order IFCs are calculated to fit the displacement-force relation~\cite{Tadano2018}. 
Note that the pseudopotentials used in both QE and VASP calculations are selected based on the best agreement between the results of the phonon dispersion under the harmonic approximation, and the results are given in Fig.~\ref{fig: harm_comp} in Appendix A. 
Based on such a foundation, we can replace the dynamical matrix data in DFPT, which is used in QE, with the one from SCPH, which is generated by VASP plus ALAMODE. 

Lastly, $T_c$ is evaluated using the MAD formula [Eq.(\ref{eq:MAD})], as explained in the previous section. 
The pseudo-Coulomb parameter $\mu^*$ is set to 0.1 to account for the screened Coulomb interaction.

\section{RESULTS}

\subsection{Electronic Structures}

\begin{figure}[t] \centering
\includegraphics[width=\linewidth]{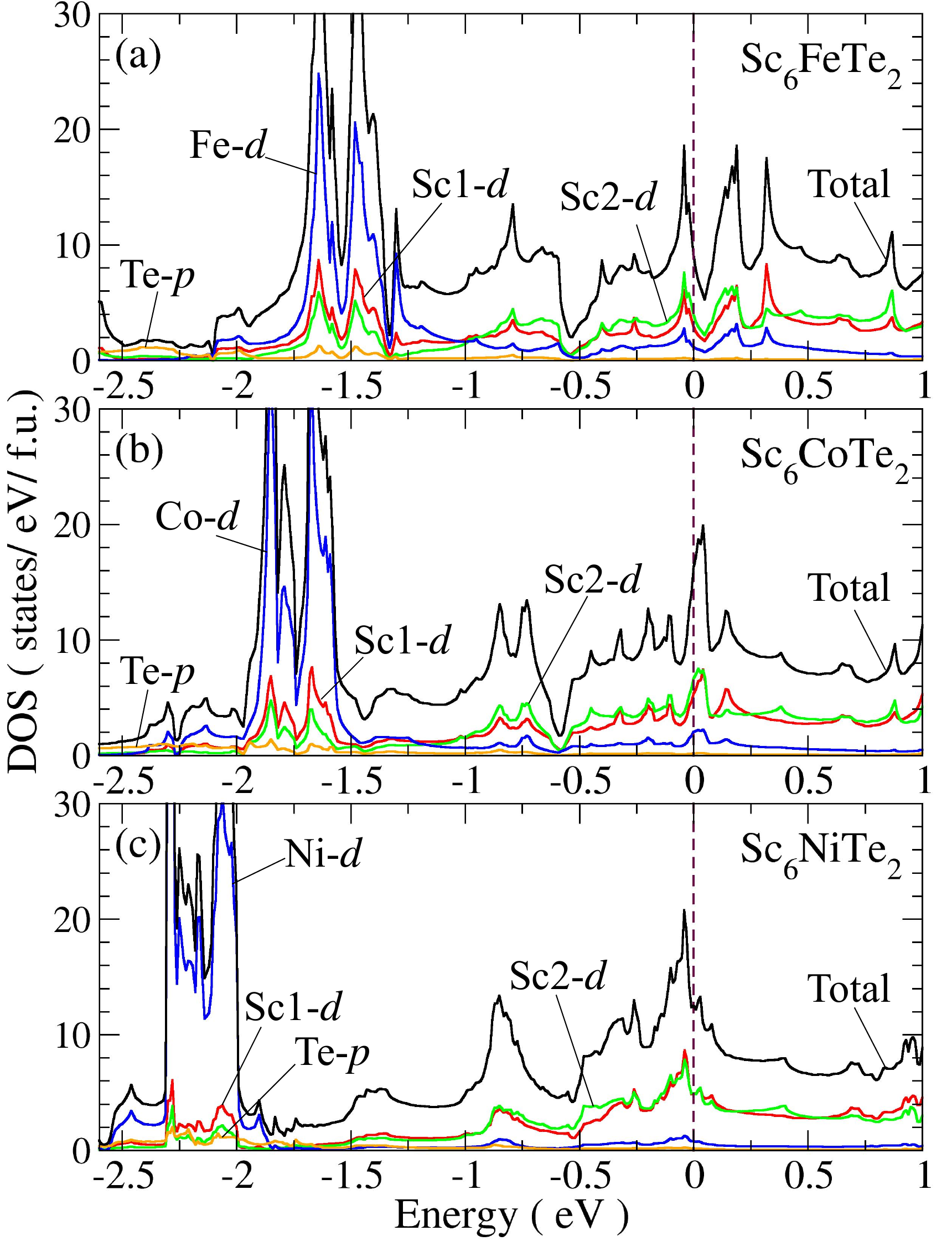}
\caption{Total and orbital-decomposed density of states (DOS) of (a) Sc$_6$FeTe$_2$, (b) Sc$_6$CoTe$_2$, and (c) Sc$_6$NiTe$_2$. The Fermi level is set to zero, as denoted by the vertical dashed line.
}
\label{fig:eldos}
\end{figure}

\begin{figure*}[t] \centering
\includegraphics[width=\textwidth]{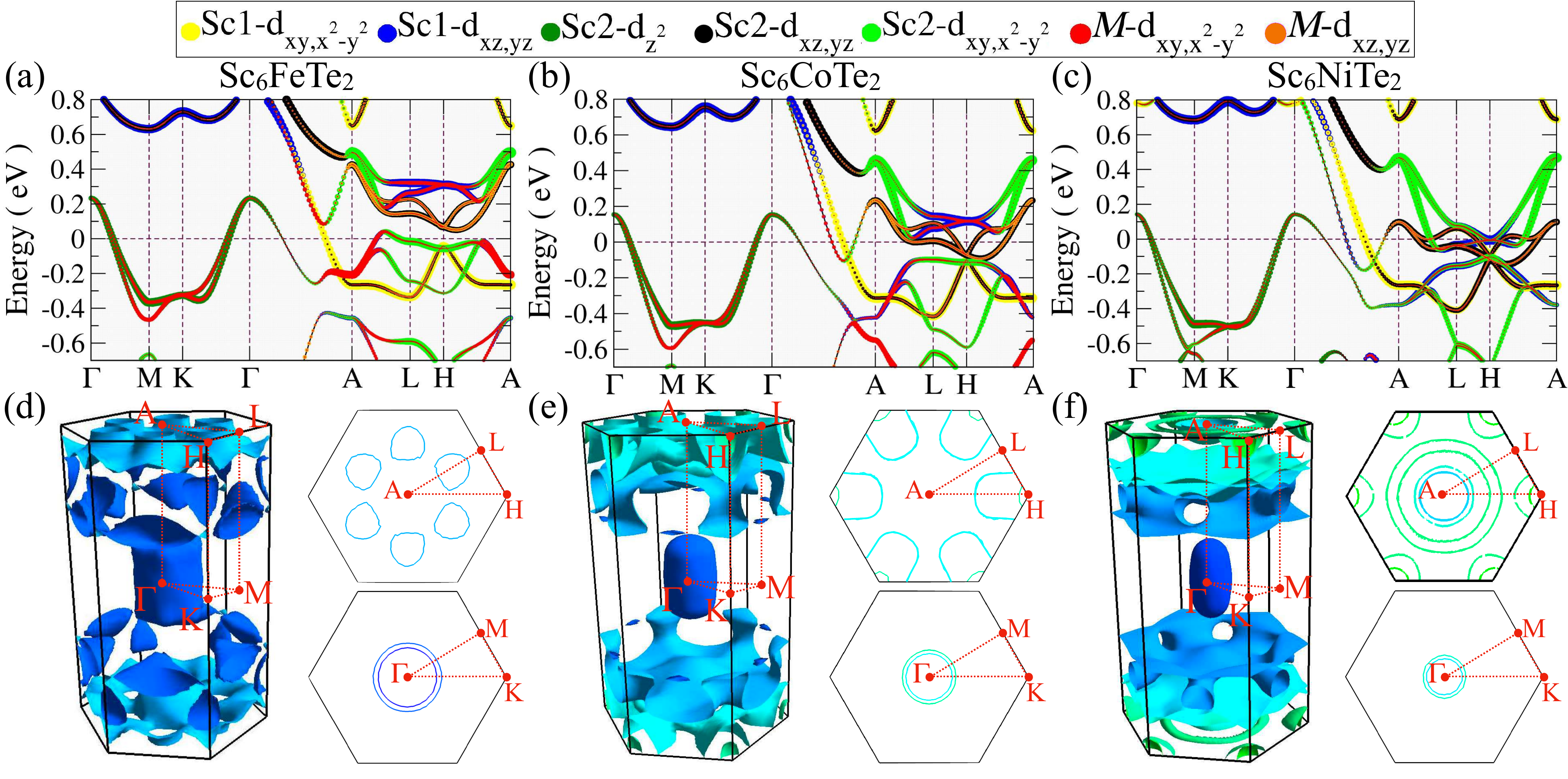}
\caption{Electronic band structures around the Fermi level (0 eV) of (a) Sc$_6$FeTe$_2$, (b) Sc$_6$CoTe$_2$, 
and (c) Sc$_6$NiTe$_2$ color-coded with the $d$-orbital projections. 
Note that the projected density of states for the $d$-orbitals not plotted contribute at least three times less to the Fermi level than those that are plotted. The Fermi surface, color-coded by band indices, and two sections along the $\Gamma$-M-K and A-L-H plane for Sc$_6M$Te$_2$ where $M=$ (d)Fe, (e)Co, and (f)Ni.}
\label{fig:elband}
\end{figure*} 

We examine the electronic structures of Sc$_6M$Te$_2$. 
Figure~\ref{fig:eldos} shows the total and orbital-decomposed DOS for Sc$_6M$Te$_2$. 
We notice that there are sharp peaks in the DOS around 1.5 eV to 2.2 eV below the Fermi level, which is dominated by the $M$-3$d$ orbitals. 
This suggests that the $M$-3$d$ orbitals are localized in space because the narrow width of the peaks and the weak hybridization with other orbitals indicates weak hopping to the Sc atoms around them. 
We see that these peaks gradually get narrower and deeper below the Fermi level as we go from Fe toward Ni.
This can be explained by the difference in the electronegativity of Fe(1.83), Co(1.88), and Ni(1.91)~\cite{Zumdahl2004} as the electrons will be more localized and more strongly bound in atoms with larger electronegativity.

Next, we discuss the electronic structures of Sc$_6M$Te$_2$ near the Fermi level.
In Fig.~\ref{fig:eldos}, we observe hybridized DOS between Sc1, Sc2, and $M$ within the energy range from -1 to 1 eV. 
Notably, there is a peak structure in the DOS at the Fermi level across all the materials, indicating large DOS at the Fermi level.  
Large DOS at the Fermi level is beneficial for phonon-mediated superconductivity~\cite{Schrieffer1971}. 

In Figs.~\ref{fig:elband}(a)-\ref{fig:elband}(c), we show the detailed orbital-projected band structures and the corresponding Fermi surface in Figs.~\ref{fig:elband}(d)-\ref{fig:elband}(f). 
In Figs.~\ref{fig:elband}(a)-\ref{fig:elband}(c), all the band structures possess a hole pocket around the $\Gamma$ point. 
This hole pocket at $\Gamma$ primarily comprises hybridized $M$-3$d_{xy,x^2-y^2}$ and Sc2-3$d_{z^2}$ orbitals.  
From Figs.~\ref{fig:elband}(d)-\ref{fig:elband}(f), we see that the hole pocket forms a cylindrical Fermi surface around $\Gamma$, and the size of this cylindrical patch decreases from $M=$ Fe to Ni, as clearly depicted in the $\Gamma$-M-K section in Figs.~\ref{fig:elband}(d)-\ref{fig:elband}(f).
We do not observe any Fermi surface pockets around the M and K points.
On the other hand, the projected band structures along A-L-H changes through electron filling. For example, from Figs.~\ref{fig:elband}(a)(b), along A-L-H, we can see that for $M=$ Fe, the $M$-3$d_{xz,yz}$and Sc2-3$d_{xz,yz}$ bands are not occupied while they are occupied in the case of $M=$ Co. 
Moreover, the Fermi surface on the A-L-H plane is more complicated than the hole pocket surrounding the $\Gamma$ point. This originates from the less dispersive bands with more crossings across the Fermi level compared to that of $\Gamma$-M-K plane, as we can see from Figs.~\ref{fig:elband}(a)-(c). The same observation can also be made by looking at the A-L-H section of the Fermi surface at Figs.~\ref{fig:elband}(d)-\ref{fig:elband}(f). 

\begin{table}[htbp]
\caption{Results of Bader charge analysis for Sc$_6M$Te$_2$ ($M$=Fe, Co, Ni). The closest ideal ionic state, as well as its oxidation state, 
is also given. The valence orbitals are $4s^23d^{1,7,8,9}$ for Sc, Fe, Co, Ni, and $5s^25p^4$ for Te. 
The Bader charge of each neutral atom is thus 3, 8, 9, 10, 6 for Sc, Fe, Co, Ni, and Te, respectively.}
\begin{ruledtabular}
\begin{tabular}{c c c }
Compound & Bader charge &  Oxidation state \\
         &  (e$^-$)& (e$^-$) \\
\hline
Sc$_6$FeTe$_2$& & \\
\hline
 Sc1& 1.95  & 2 (Sc$^{1+}$)\\
 Sc2& 1.88  & 2 (Sc$^{1+}$)\\
 Fe & 10.97 & 11 (Fe$^{3-}$)\\
 Te & 7.74  & 8 (Te$^{2-}$) \\
 \hline
Sc$_6$CoTe$_2$ & &\\ 
\hline
 Sc1& 1.94  & 2 (Sc$^{1+}$)\\
 Sc2& 1.91  & 2 (Sc$^{1+}$)\\
 Co & 11.92 & 12 (Co$^{3-}$)\\
 Te & 7.76  &  8 (Te$^{2-}$) \\
 \hline
 Sc$_6$NiTe$_2$& &\\
\hline
 Sc1& 1.96  & 2 (Sc$^{1+}$)\\
 Sc2& 1.94  & 2 (Sc$^{1+}$)\\
 Ni & 12.69 & 12 (Ni$^{2-}$)\\
 Te & 7.78  &  8 (Te$^{2-}$) \\
 
\end{tabular}
\end{ruledtabular}
\label{table:Bader}
\end{table}
Last, we discuss the absence of magnetism in Sc$_6M$Te$_2$ ($M=$ Fe, Co, and Ni).
In the experiment, no sign of magnetism has been reported in these materials~\cite{Maggard2000,Shinoda2023}, which is essential for the emergence of superconductivity but quite nontrivial because they include magnetic elements $M=$ Fe, Co, and Ni.
In fact, no magnetic states can be stabilized for $M$= Co, Ni in our spin-polarized self-consistent DFT calculations. 
For $M=$ Fe, we obtain both a ferromagnetic and a nonmagnetic solution. Although the energy difference between them is just $\sim$1 meV, the total energy of the nonmagnetic solution is lower in our calculations. This result is consistent with the experimental findings~\cite{Shinoda2023}. To investigate the underlying reason, we perform the Bader charge analysis, which is a useful charge counting analysis done by segregating atoms along the minimum in the electron density~\cite{Tang2009}. The Bader charges, which are the counted valence electrons, and the corresponding nearest oxidation state are given in Table~\ref{table:Bader}. 
Interestingly, every Sc atom donates roughly one electron to the neighboring $M$ and Te, transferring six electrons. 
The pair of Te atoms would share three electrons and $M$ would take the remaining three electrons. 
Consequently, the 3$d$ electron shell of Co and Ni is fully occupied. 
This explains why the potential magnetism brought by Fe, Co, and Ni does not show up in the Sc$_6M$Te$_2$ systems. 
Note that this electron transfer occurs due to the difference in electronegativity where Sc, Fe, Co, Ni, and Te are of value 1.36, 1.83, 1.88, and 1.91, 2.10, respectively~\cite{Zumdahl2004}. 
Crucially, this electron reallocation prevents the potentially detrimental effect on superconductivity caused by the partially filled 3$d$ shell. 
Moreover, since the Fe 3$d$ shell is not fully filled, we expect a more unstable nonmagnetic state than the rest, resonating with the absence of stable ferromagnetic results in DFT self-consistent calculation for $M=$ Co and Ni.   

\begin{figure*}[t] \centering
\includegraphics[width=\textwidth]{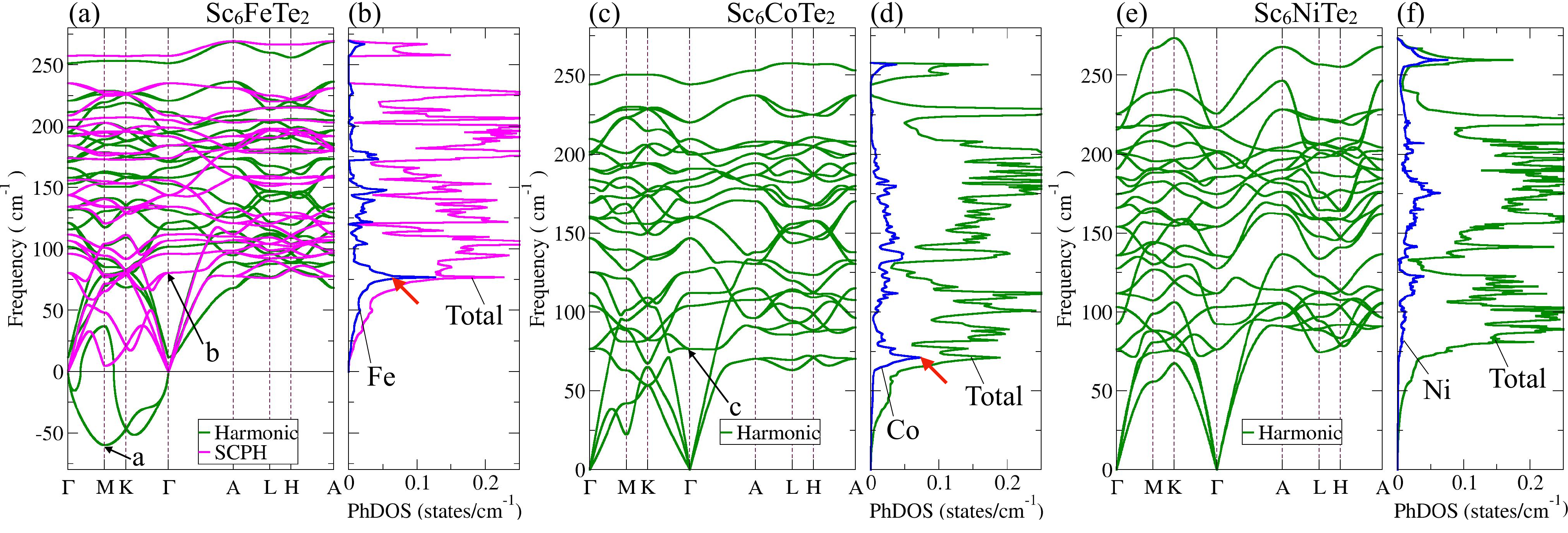}
\caption{Phonon band structures, total and $M$-projected phonon density of states (PhDOS) of (a)(b) Sc$_6$FeTe$_2$, (c)(d) Sc$_6$CoTe$_2$, and (e)(f) Sc$_6$NiTe$_2$. The magenta curves in (a) and (b) are the anharmonicity-renormalized phonon bands. The displacement pattern of the phonon bands indicated by markers ``a", ``b", and ``c" are given in Fig.~5. Also, the red arrows point to the localized phonon bands arising from the Fe and Co atoms.}
\label{fig: ph}
\end{figure*}

\subsection{Phonon Properties}
\label{subsection:Phonon_properties}

\begin{figure}[H] \centering 
\includegraphics[width=\columnwidth]{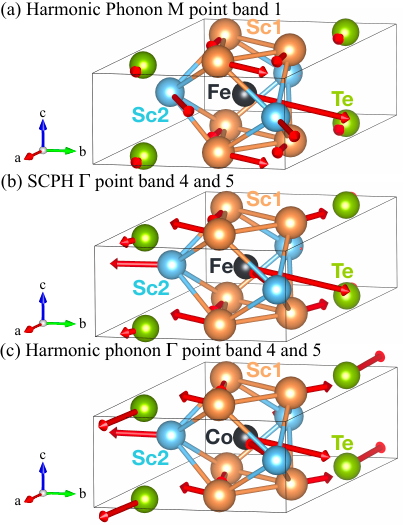}
\caption{Displacement pattern of (a) the imaginary frequency phonon mode at the M point of Sc$_6$FeTe$_2$ [marker ``a" in Figure 4(a)] (b) the doubly-degenerated phonon bands number 4 and 5 around 75 cm$^{-1}$ at the $\Gamma$ point of Sc$_6$FeTe$_2$ [marker ``b" in Figure 4(a)] and (c) Sc$_6$CoTe$_2$ [marker ``c" in Figure 4(c)]. All displacement modes shown here are in-plane.}
\label{fig: displacepat}
\end{figure}

Now, we discuss the phonon properties of the Sc$_6M$Te$_2$ systems. 
The calculated phonon bands are given in Fig.~\ref{fig: ph}(a)(c)(e). 
As shown in Fig.~\ref{fig: ph}(a), under the harmonic approximation, imaginary frequency phonon bands exist along the $\Gamma$-M-K plane of Sc$_6$FeTe$_2$. 
We mark the imaginary frequency phonon bands at the M point as ``a" in Fig.~\ref{fig: ph}(a).
The displacement amplitude along the polarization vector of such band is displayed in Fig.~\ref{fig: displacepat}(a).
We see that the origin of such an imaginary frequency mode is the in-plane vibration of the Fe atoms.
On the other hand, Sc$_6$CoTe$_2$ and Sc$_6$NiTe$_2$ do not have imaginary frequencies under the harmonic approximation, as shown in Fig.~\ref{fig: ph}(c) and (e).
This suggests that Fe atoms in Sc$_6$FeTe$_2$ will show larger atomic displacements in the in-plane direction and manifest stronger anharmonicity compared to $M$=Co and Ni cases. Note that anharmonicity refers to the deviation of the potential energy surface from the parabolic behavior near the reference structure.

However, the instability is overestimated in the DFT calculation because no structural phase transitions have been observed in the experiment~\cite{Shinoda2023}. Therefore, we use the SCPH theory to obtain a positive dispersion in a physically reasonable way, as we discussed in Sec.~\ref{subsec:theory_SCPH}.
The obtained SCPH dispersion with $T_{\rm SCPH}$ = 20 K is displayed with magenta lines in Fig.~\ref{fig: ph}(a). 
Comparing the harmonic and SCPH bands, the higher-frequency portion of the phonon bands remains unchanged, while the imaginary phonon modes are stabilized, resulting in dispersive phonon bands in the low-frequency region. 

\begin{figure*}[t] \centering 
\includegraphics[width=\textwidth]{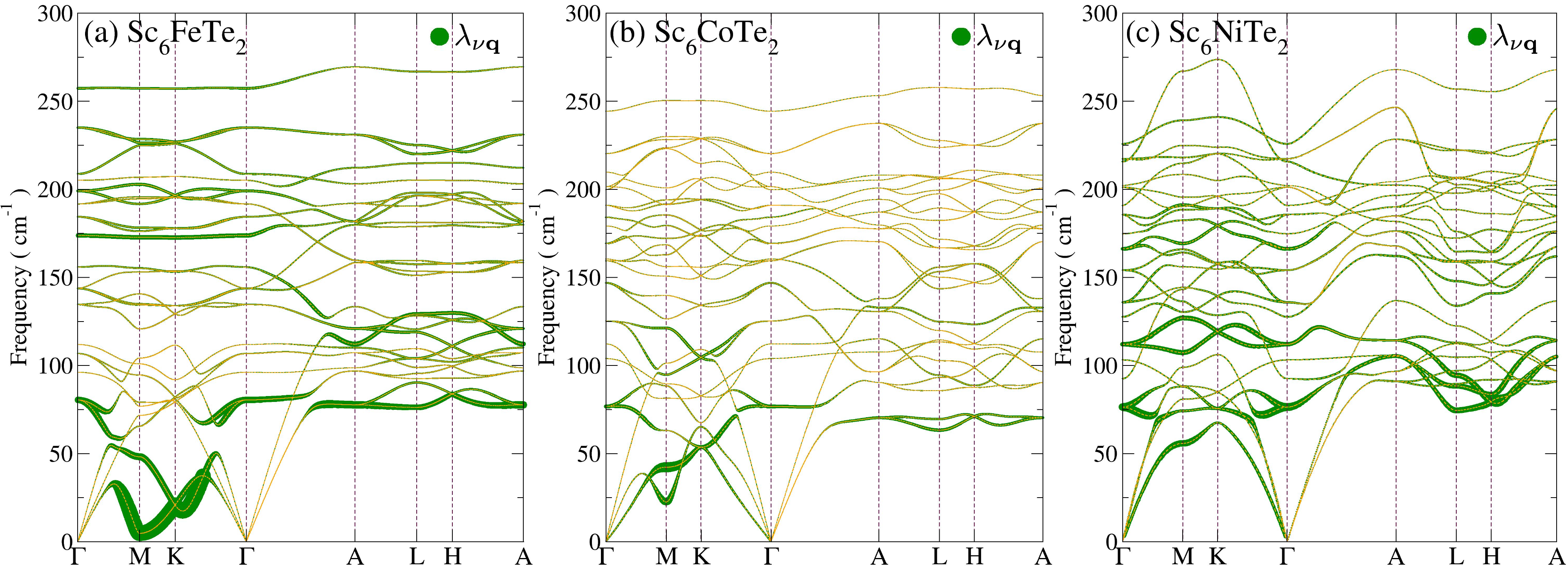}
\caption{Phonon band structure color-coded with the relative size of electron-phonon coupling strength $\lambda_{\nu\textbf{q}}$ 
for Sc$_6M$Te$_2$. Orange lines represent the phonon bands. Note that the thickness of the line is normalized to $\lambda_{\nu\textbf{q}}$ of each material.
}
\label{fig: ph_epc}
\end{figure*}

In Figs.~\ref{fig: ph}(a)(c)(e), we can see common features across $M=$ Fe, Co, and Ni. 
All spectral widths are roughly 250 cm$^{-1}$.
Also, the phonon band structures of Sc$_6M$Te$_2$ are generally dispersed in the low-frequency region and flatter in higher-frequency regions. 
For Sc$_6$NiTe$_2$, the phonon bands are more dispersive than the others.
The similarity of the phonon band structures can also be well observed by the total phonon density of states (PhDOS), as given in Figs.~\ref{fig: ph}(b)(d)(f). We can see that the peak positions and magnitudes of the total PhDOS are qualitatively similar for $M=$ Fe, Co, and Ni. Interestingly, the difference lies in the $M$ atom-projected PhDOS. Compare Figs.~\ref{fig: ph}(b)(d)(f), as indicated by the red arrows, around 75 cm$^{-1}$, only the renormalized phonon bands of Sc$_6$FeTe$_2$ and the harmonic phonon bands of Sc$_6$CoTe$_2$ have a PhDOS peak. This indicates the emergence of localized $M$ phonon modes for only Sc$_6$FeTe$_2$ and Sc$_6$CoTe$_2$. 
Thus, we take the phonon bands around 75 cm$^{-1}$ at the $\Gamma$ point as representatives to investigate their vibration patterns. They are degenerated phonon bands with index number 4 and 5 for both materials and we mark them as ``b" in Fig.~\ref{fig: ph}(a) and ``c" in Fig.~\ref{fig: ph}(c). 
In Figs.~\ref{fig: displacepat}(b)(c), we see that these modes originate from in-plane vibrations of $M$.
On the contrary, the Ni PhDOS of Sc$_6$NiTe$_2$ is fairly uniform throughout the spectrum, as shown in Fig.~\ref{fig: ph}(f).

Significantly, we realize that such a difference in the distribution of the $M$-projected PhDOS is related to the EPC strength. 
Figure~\ref{fig: ph_epc} displays the phonon bands of Sc$_6M$Te$_2$ with indication of the relative magnitude of the EPC strength $\lambda_{\nu \textbf{q}}$ [Eq.~(\ref{eq:lambda})]. 
Note that a hot spot of $\lambda_{\nu \textbf{q}}$ indicates a strong pairing interaction between electrons at \textbf{k} and \textbf{k}+\textbf{q} mediated by the $\nu$ phonon mode .
For $M=$ Fe and Co, Sc$_6M$Te$_2$ has large $\lambda_{\nu \textbf{q}}$ in phonon modes with frequencies of $\sim$25-40 cm$^{-1}$ around M and K. 
For $M=$ Fe,  as is obvious from Fig.~\ref{fig: ph}(a), the anharmonic effect is significant in this segment of the phonon bands.
The second hot spot lies around 75 cm$^{-1}$. This segment highlights the difference in the $M$ atom-projected PhDOS as observed through Figs.~\ref{fig: ph}(b)(d)(f). If we look at Fig.~\ref{fig: ph_epc} (c), we would see that $\lambda_{\nu \textbf{q}}$ of Sc$_6$NiTe$_2$ is uniformly distributed with respect to phonon momentum and frequency.
From Fig.~\ref{fig: ph_epc}, we reveal the significance of these low-frequency phonons in understanding the chemical trend of the EPC and consequently $T_c$ for Sc$_6M$Te$_2$ ($M=$ Fe, Co, Ni).
In Sec. IV A, we will further discuss that we could interpret these low-frequency phonons that generate strong EPC as ``rattling" phonons. 

\subsection{Eliashberg function and Superconductivity}

\begin{figure*}[tbph] \centering
\includegraphics[width=0.8\textwidth]{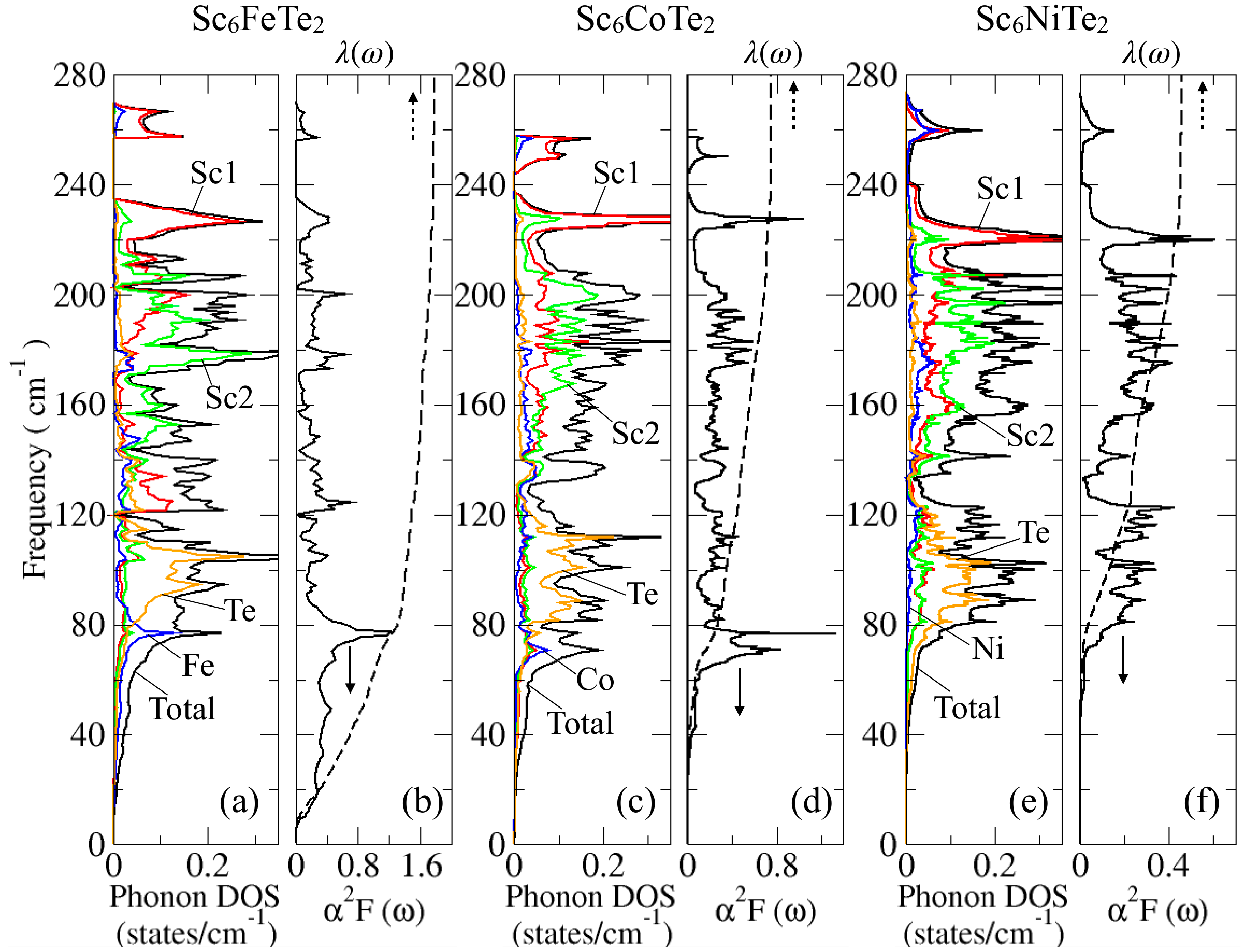}
\caption{Atom-decomposed phonon density of states (Phonon DOS), as well as the Eliashberg spectral functions $\alpha^2F (\omega)$ and the frequency-resolved electron-phonon coupling strength $\lambda(\omega)$ of (a)(b) Sc$_6$FeTe$_2$, (c)(d)Sc$_6$CoTe$_2$, and (e)(f) Sc$_6$NiTe$_2$. Note that $\lambda(\omega)=2\int^{\omega}_0\frac{\alpha^2F(\omega')}{\omega'}d\omega'$ and thus the electron-phonon coupling constant $\lambda$ is the value obtained after integrating over the whole phonon spectrum.}
\label{fig: elph}
\end{figure*}

\begin{table}[htbp]
\caption{Calculated electron-phonon coupling constant $\lambda$, logarithmic average phonon frequency $\omega_{\rm log}$, total [$N(\epsilon_\text{F})$] and $M$-projected [$N_{M}(\epsilon_\text{F})$] density of states at the Fermi level, and the superconducting transition temperature ($T_c$) of Sc$_6M$Te$_2$ ($M$=Fe, Co, Ni). The last column includes the temperature ($T_{\rm SCPH}$) used in the SCPH theory. The smearing parameter is 0.02 eV, and the pseudo-Coulomb parameter $\mu^*$ is 0.1.}
\begin{ruledtabular}
\begin{tabular}{c c c c c c c}
System & $\lambda$ &  $\omega_{\rm log}$  & $N(\epsilon_\text{F})$& $N_{M}(\epsilon_\text{F})$& $T_c$ (Exp.$^a$)  &   $T_{\rm SCPH}$\\
$M$  &     &  (K)  & (1/eV/f.u.)& (1/eV/$M$) &  (K)  & (K) \\
\hline
Fe& 1.78& 72& 9.04 & 1.16& 9.4 (4.7)& 20  \\
Fe& 1.67& 79& 9.04 & 1.16& 9.9 (4.7)& 50  \\
Fe& 1.46& 95& 9.04 & 1.16& 10.5 (4.7)& 100  \\
Co& 0.75& 152& 16.30 & 2.03& 6.2 (3.5)& -  \\
Ni& 0.46& 187& 12.46 & 0.79& 1.6 (2.7)& -  \\
\end{tabular}
\end{ruledtabular}
{$^a$Experimental data taken from Ref.~\cite{Shinoda2023}}
\label{table:eliash}
\end{table}
In Figs.~\ref{fig: elph}(b), \ref{fig: elph}(d) and \ref{fig: elph}(f), 
the calculated Eliashberg spectral function ($\alpha^2F(\omega)$) is depicted alongside the PhDOS. 
$\alpha^2F(\omega)$ provides insight into the EPC strength ($\lambda$). 
Typically, peaks in $\alpha^2F(\omega)$ align with those in PhDOS. 
However, a significant $\alpha^2F(\omega)$ contribution to $\lambda$ emerges in the 60 to 80 cm$^{-1}$ range for Sc$_6$FeTe$_2$ and Sc$_6$CoTe$_2$, which is absent in Sc$_6$NiTe$_2$. 
This distinction can be attributed to the presence or absence of the $M$-dominated low-frequency phonons discussed in the previous subsection.
For Sc$_6$FeTe$_2$, $\alpha^2F(\omega)$ manifests a wide plateau in 20 to 50 cm$^{-1}$ frequency range, yielding significant contributions to $\lambda$. 
At higher frequency regions above 80 cm$^{-1}$, the EPC contribution coincides with the peaks of PhDOS, initially from the Te phonon bands followed by the Sc phonon bands. 
Figure~\ref{fig: elph} illustrates, firstly, the critical role of the $M$ atom in the EPC of Sc$_6M$Te$_2$. Second, it shows a correlation between low-frequency vibrations of $M$ and enhanced EPC. Finally, it demonstrates the dominance of the anharmonicity-driven EPC for $M=$ Fe.

By integrating $\alpha^2F(\omega)/\omega$ over frequency, we obtain the EPC constant $\lambda$ [Eq.~(\ref{eq:lama2F})].
As shown in Table~\ref{table:eliash}, the resultant $\lambda$ values are 1.78, 0.75, and 0.46 for $M=$ Fe, Co, and Ni, respectively. 
Utilizing the calculated $\lambda$, we predict $T_c$ values using the MAD formula [Eq.~(\ref{eq:MAD})]~\cite{McMillan1968,Dynes1972,Allen1975}. 
The pseudo-Coulomb interaction ($\mu^*$) is treated as an empirical parameter and is set to 0.1.
The predicted $T_c$ values for Sc$_6M$Te$_2$ are 9.4 K, 6.2 K, and 1.6 K for $M=$ Fe, Co, and Ni, respectively. 
The calculated values of $T_c$ align with the observed chemical trends~\cite{Shinoda2023}, with $M=$ Fe ranking the highest and $M=$ Ni the lowest. 
Also, the predicted values of $T_c$ are comparable to the measured values. 
Thus, we have revealed that the Sc$_6M$Te$_2$ ($M=$ Fe, Co, Ni) are phonon-mediated superconductors.
Note that the calculated $\lambda$ and $T_c$ of Sc$_6$FeTe$_2$ is little affected by $T_{\rm SCPH}$ as shown in Table~\ref{table:eliash}.
While we use the calculation of $T_{\rm SCPH}=$ 20 K as the representative, the discussion remains robustly for a wide range of $T_{\rm SCPH}$. See Appendix B for further discussions on $T_{\rm SCPH}$ of Sc$_6$FeTe$_2$.

\begin{figure*}[t] \centering
\includegraphics[width=0.78\textwidth]{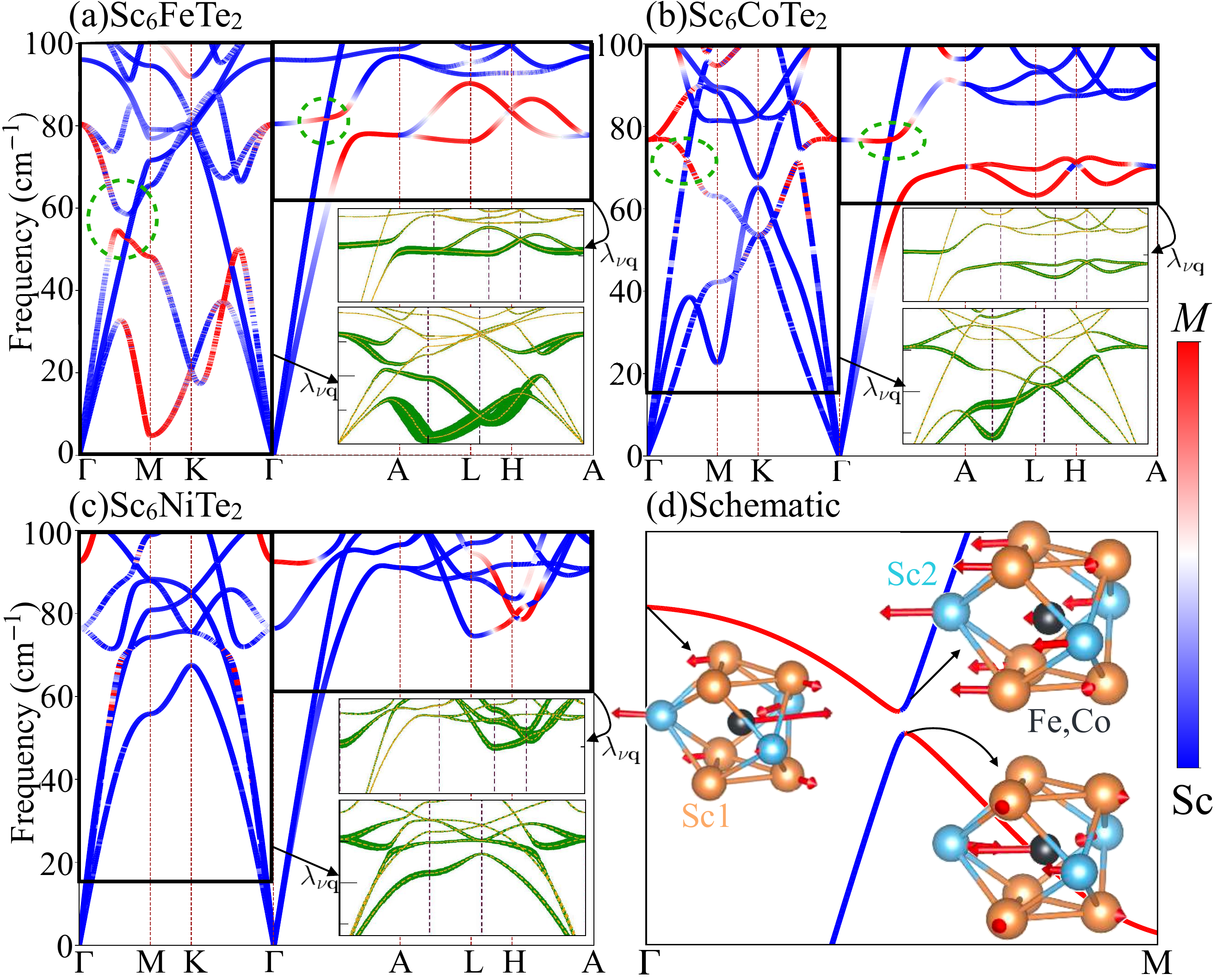}
\caption{Rattling phonon modes of Sc$_6$FeTe$_2$ and Sc$_6$CoTe$_2$. Phonon band structures in low-frequency regions are color-coded based on their relative magnitude of atomic displacement for the rattler ($M$ atom) and the cage (Sc atoms) for $M=$ Fe(a), Co(b), and Ni(c). Insets of each are the phonon band structures in low-frequency regions color-coded with the relative size of the electron-phonon coupling strength $\lambda_{\nu\textbf{q}}$. (d) Schematic of the avoid crossing between the rattling mode and longitudinal acoustic mode as well as the displacement pattern around the crossing for $M=$ Fe and Co.}
\label{fig: ph-rat}
\end{figure*}

Finally, we discuss the possible origins of the deviation between the calculated and experimental $T_c$. One possibility lies in the influence of spin fluctuation coming from the $d$-electrons in the system. Kawamura \textit{et al.}~\cite{Kawamura2020} carried out systematic {\it ab initio} calculations including both EPC and spin fluctuation based on density functional theory for superconductors (SCDFT) for all the elemental metals. Their SCDFT calculations showed that for some elemental metals, including spin fluctuation could significantly reduce the phonon-mediated $T_c$. In particular, they reported that elemental Sc metal has an EPC constant $\lambda$ of 0.398 and a $T_c$ of 2.7 K when the spin fluctuation is not considered. However, when the spin fluctuation is taken into account, the superconductivity disappears, agreeing with the experiments. We notice that the DOS at the Fermi level in hexagonal Sc is rather large, being 2.01 states/eV/Sc atom~\cite{Kawamura2020}, and thus, spin fluctuation would be quite significant. Figure 2 shows that the projected DOS for Sc at the Fermi level in Sc$_6M$Te$_2$ is also rather large, being 1.65, 2.06, and 1.38 states/eV/Sc atom, respectively, for $M=$  Fe, Co, and Ni. Since two-thirds of atoms in Sc$_6M$Te$_2$ ($M$ = Fe, Co, Ni) are Sc, the spin fluctuation in Sc$_6M$Te$_2$ may affect the predicted phonon-mediated $T_c$ in Sc$_6M$Te$_2$. In particular, this perhaps explains why the calculated $T_c$ values in $M=$ Fe and Co are higher than the experimental values (Table III).

Also, such deviation of $T_c$ could arise from the isotropic gap function that the MAD formula assumes. Given the complicated Fermi surface of Sc$_6M$Te$_2$, it is an interesting future problem of how much the predicted $T_c$ is affected if we go beyond the MAD formula.

In conclusion, our findings suggest that superconductivity in Sc$_6M$Te$_2$ ($M=$ Fe, Co, Ni) is phonon-mediated, where low-frequency $M$ vibrations and anharmonic phonon modes play a crucial role. Also, our study suggests by utilizing the electronegativity difference, we can manipulate the electron fillings of the $d$-orbitals to make the combination of a non-superconducting elemental metal, Sc in the present case, and magnetic element become superconducting. In general, this proposition could serve as a potential explanation or direction of investigation for Sc-included systems, including one recent report on the discovery of superconductivity of ScPdGe and ScPdSi~\cite{Shinoda2024}.

\section{Discussions}

In this section, we further extend the discussions toward the rattling phonons and the magnetic suppression of Sc$_6$MnTe$_2$. 

\subsection{Rattling phonons}
From our results (Figs.~\ref{fig: ph},~\ref{fig: ph_epc}, and~\ref{fig: elph}), we show that low-frequency phonons from $M$ are the key factor to understand the trend in $T_c$ for Sc$_6$MTe$_2$. Also, note that in Fig.~\ref{fig: displacepat}, we learn that these phonons are mainly single-atom vibrations. These results prompt us to interpret this key factor to be ``rattling phonons", where atoms ``rattle" alone in the system, producing low-frequency vibrational modes and large EPC~\cite{Hiroi2007,Nagao2009,Winiarski2016}.   

Here, we discuss the phonon dispersion of a system with rattling phonons, which in our case are the $M=$ Fe and Co phonon modes. Christensen \textit{et al.}~\cite{Christensen2008} showed by spring model calculations and inelastic neutron scattering experiments that the hallmark of a flat rattling mode is its avoidance of crossing with the longitudinal acoustic mode~\cite{Christensen2008}.
In Figs.~\ref{fig: ph-rat}(a)-\ref{fig: ph-rat}(c), we present the low-frequency segment of the phonon band structures.
Notably, in both Sc$_6$FeTe$_2$ and Sc$_6$CoTe$_2$, the flat phonon band around 75 cm$^{-1}$ predominantly reflects the $M$ vibration mode. It persists over a wide range along the high symmetry line, particularly the A-L-H plane. Furthermore, as circled in green in Figs.~\ref{fig: ph-rat}(a)and(b), we can discern a gap between this flat phonon band and the longitudinal acoustic mode along the $\Gamma$-M and $\Gamma$-A line, displaying the hallmark of rattling modes~\cite{Christensen2008}. 
Fig.~\ref{fig: ph-rat}(d) shows the schematic of such avoided crossing and the displacement pattern of $M=$ Fe and Co around the gap. 
We notice the displacement of the Sc cage to dominate along the acoustic mode before and after the gap. Moreover, we notice the dominance of the rattler displacements from Fe and Co to persist on the rattling phonon band.
On the contrary, from Fig.~\ref{fig: ph-rat}(c), we notice the absence of such a feature in Sc$_6$NiTe$_2$. This is consistent with the PhDOS shown in Figs.~\ref{fig: ph}. The insets of Fig.~\ref{fig: ph-rat}(a)-(c) are the low-frequency segment of Fig.~\ref{fig: ph_epc}, we clearly see that the EPC of Sc$_6$FeTe$_2$ and Sc$_6$CoTe$_2$ is highly concentrated in the region of the rattling phonons. This suggests that Sc$_6M$Te$_2$ is a case to demonstrate the EPC driven by these localized rattling modes that can be further tuned by the degree of lattice anharmonicity.

\subsection{Magnetic suppression of Sc$_6$MnTe$_2$}

Now let us inquire further into the absence of the magnetic moment on $M$ in Sc$_6M$Te$_2$~\cite{Shinoda2023}. 
As elaborated in the last part of Sec. III A, our Bader charge analysis reveals an electron transfer from Sc to $M$ because Sc displays the lowest electronegativity among the constituent atoms. 
This electron transfer provides enough electrons to saturate the 3$d$-orbitals of Fe, Co, and Ni, 
thus suppressing the formation of the magnetic moment on $M$. 
In the experiments by Shinoda \textit{et al.}~\cite{Shinoda2023}, superconductivity is not observed for $M=$ Mn. 
Since Mn has a much lower electronegativity of 1.55 compared with that of Fe, Co, and Ni~\cite{Zumdahl2004},
the large electron transfer from Sc to $M$ mentioned above would be weaker; thus, Sc$_6$MnTe$_2$ may have a magnetic ground state. 
This suggests that superconductivity in Sc$_6$MnTe$_2$ is perhaps suppressed by magnetic pair breaking~\cite{Guo1990} due to the magnetic moments on the Mn atoms. 
To verify this hypothesis, we further perform self-consistent spin-polarized electronic
structure calculations for Sc$_6$MnTe$_2$. Indeed, we find that Sc$_6$MnTe$_2$ has a stable ferromagnetic state with a local magnetic moment of 1.4 $\mu_B$ on Mn.
The total energy of the ferromagnetic state is 25 meV/f.u. lower than the nonmagnetic state.
Moreover, we perform the Bader charge analysis toward Sc$_6$MnTe$_2$ with the final results giving us 1.98, 10.01, and 7.71 e$^-$ for the Sc, Mn, and Te atoms, respectively. The Bader charges correspond to the nearest oxidation state of 2 e$^-$ (Sc$^{1+}$), 10 e$^-$ (Mn$^{3-}=$ Ni valence state $4s^23d^8$), and 8 e$^-$ (Te$^{2-}$). 

\section{CONCLUSIONS}

In conclusion, we have carried out a systematic \textit{ab initio} computational study on the magnetic, electronic, phonon, and superconducting properties of newly reported transition-metal-based superconductors Sc$_6M$Te$_2$ ($M$ = Fe, Co, Ni)~\cite{Shinoda2023} by using theoretical methods of DFT, DFPT and SCPH.
First, our DFT calculations reveal that Sc$_6M$Te$_2$ ($M$ = Fe, Co, Ni) have nonmagnetic metallic ground states with multiple Fermi surface pockets.
The Bader charge analysis indicates a significant electron transfer from Sc to $M$ (where $M$ = Fe, Co, Ni) due to the substantial electronegativity difference. 
This charge transfer fills the $M$ 3$d$ orbitals, thus suppressing the magnetism from the $M$ atoms. 
Second, combining DFPT and SCPH, we obtain the renormalized phonon bands for the anharmonic Sc$_6$FeTe$_2$ as we observe flat phonon bands near 75 cm$^{-1}$ and dispersive low-frequency phonon bands emerging after considering the anharmonic effect.
Interestingly, these two emerging features are rattling phonon bands holding large EPC at low-frequency regions in Sc$_6M$Te$_2$ when $M =$ Fe and Co. 
Quantitatively, they also give rise to prominent peaks in the calculated Eliashberg spectral function. 
These rattling mode-enhanced EPCs lead to a rather high superconducting transition temperature ($T_c$) of $\sim$10 K for $M=$ Fe.
Our DFPT calculations also predict Sc$_6M$Te$_2$ ($M=$ Co and Ni) to be phonon-mediated superconductors with lower $T_c$ values of 6.2 K and 1.6 K, respectively. 
The chemical trend of the predicted $T_c$ for Sc$_6M$Te$_2$ from $M=$ Fe to Co and Ni agrees well with the experimental observation~\cite{Shinoda2023}. 
Importantly, we attribute such a chemical trend to the phonon bands where we observe the progression of increasing rattling behavior from $M=$ Ni to Co and then the increasing anharmonicity from $M=$ Co to Fe.  
In short, our {\it ab initio} study not only shows that the superconductivity observed in 3$d$ transition metal compounds Sc$_6M$Te$_2$ ($M =$ Fe, Co, and Ni) is phonon-mediated but also suggests a way to design transition-metal-based superconductors by combining non-superconducting and magnetic metal elements with different electronegativities.

\section*{APPENDIX A: Comparison of the harmonic phonons using DFPT and the frozen phonon method}

\begin{figure}[h] \centering
\includegraphics[width=\columnwidth]{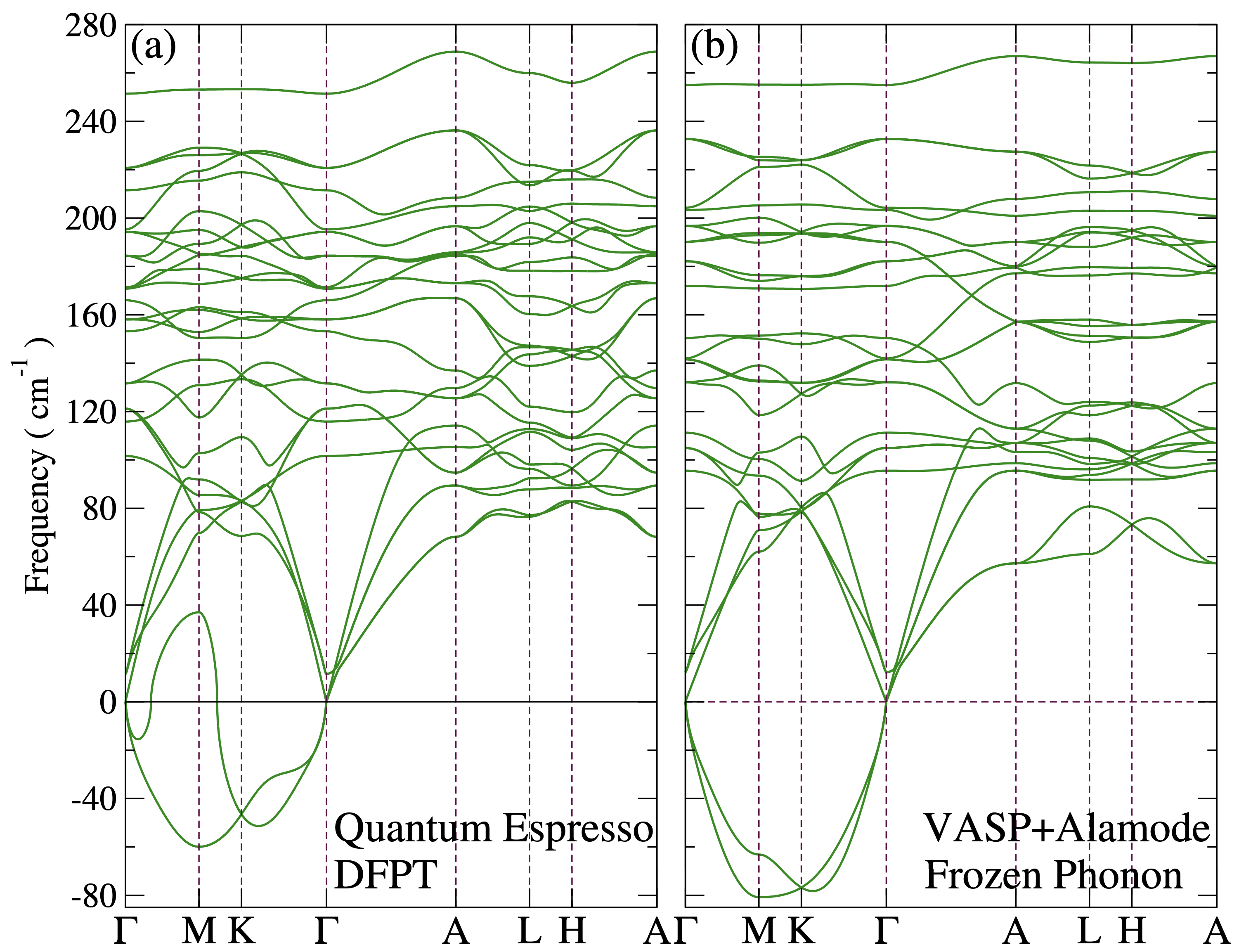}
\caption{Phonon band structures calculated under the harmonic approximation using (a) DFPT in the QE package 
and (b) frozen phonon method in the VASP plus ALAMODE package.}
\label{fig: harm_comp}
\end{figure}

In Sec. II, we mentioned replacing the harmonic phonon bands with the renormalized phonon bands 
from the SCPH theory to evaluate the EPC matrix element. 
Eq.~(\ref{eq:SCPH}) shows that the fourth-order IFCs are corrected on the harmonic phonon frequency. 
Thus, to merge the information and calculate the EPC matrix element [Eq.~(\ref{eq:g})], 
we need to double-check the agreement among the harmonic phonons. Figure~\ref{fig: harm_comp} displays the dispersion relations calculated using DFPT and the frozen phonon method. 
Clearly, Fig. \ref{fig: harm_comp} shows that the two sets of dispersion relations agree in the frequency range above 80 cm$^{-1}$.
Also, the imaginary frequency modes on the $\Gamma$-M-K plane appear in both cases.
This is thus an appropriate foundation for further calculating the EPC using 
the renormalized phonon frequencies and eigenvectors from the SCPH theory.   

\section*{APPENDIX B: Effect of $T_{\rm SCPH}$ on phonon and electron-phonon properties}

In this Appendix, we examine the effect of $T_{\rm SCPH}$ on the phonon properties and EPC. 
As mentioned in Sec. II B, a $T_{\rm SCPH}$ of 20 K was used in the Bose-Einstein distribution of the self-consistent SCPH equation [Eq.~(\ref{eq:SCPH})]. 
We choose the temperature at which the phonon bands are stable but softer than Sc$_6$CoTe$_2$, reflecting the stronger anharmonicity in Sc$_6$FeTe$_2$. 
We consider that the treatment is valid because no structural phase transition has been experimentally observed for Sc$_6$FeTe$_2$. 
Furthermore, as Table~\ref{table:eliash} shows, $T_c$ remains almost the same when $T_{\rm SCPH}$ increases. 
This indicates that the choice of $T_{\rm SCPH} =$ 20 K is appropriate.

\begin{figure}[tbph] \centering
\includegraphics[width=\columnwidth]{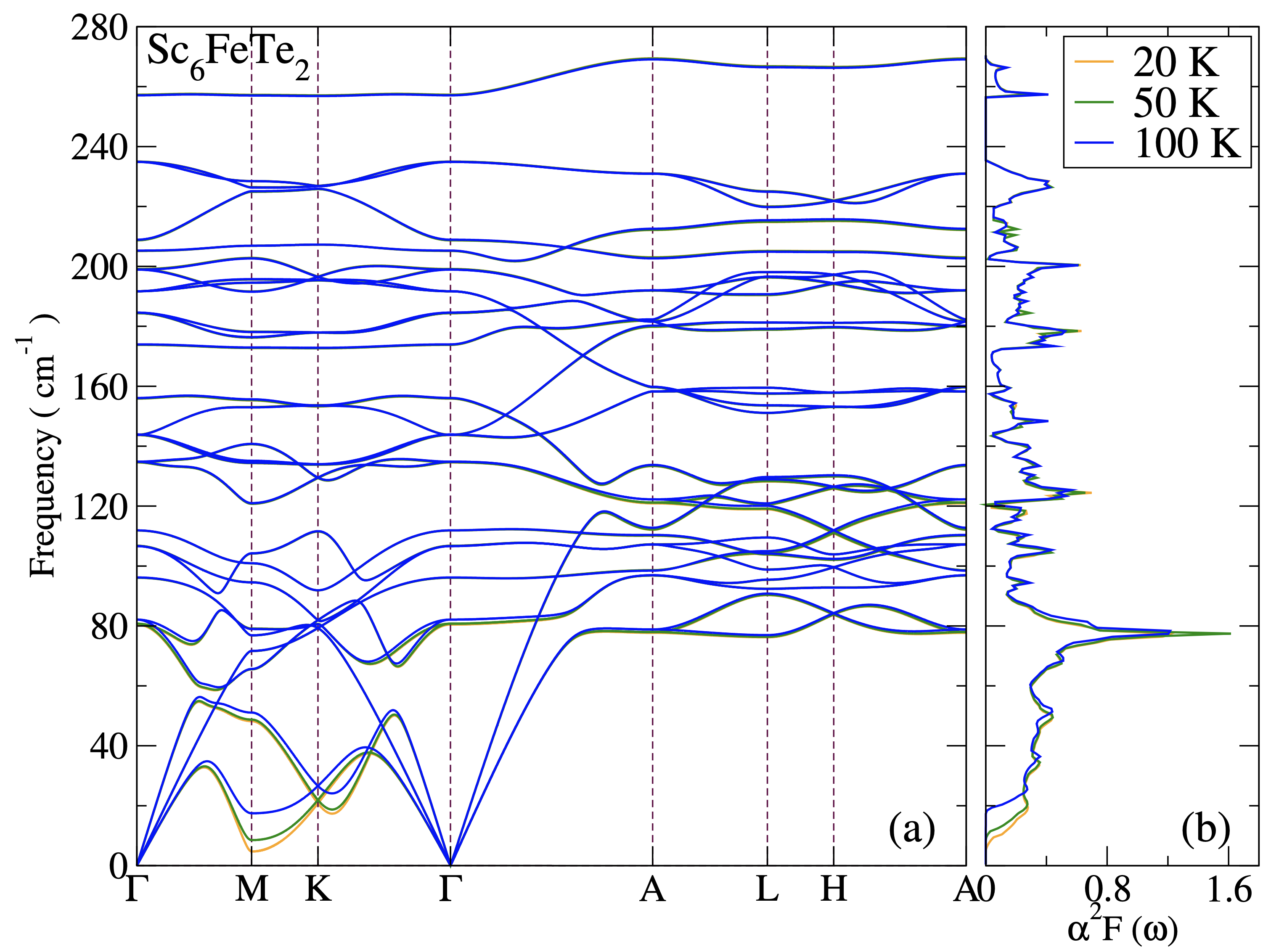}
\caption{(a)Phonon band structures and (b) Eliashberg spectral functions $\alpha^2 F$ of Sc$_6$FeTe$_2$ calculated using different $T_{\rm SCPH}$ values in the SCPH treatment of anharmonicity.}
\label{fig: t_scph}
\end{figure}

Nevertheless, we do notice that $\lambda$ drops and $\omega_{\rm log}$ rises with increasing $T_{\rm SCPH}$ (Table~\ref{table:eliash}). 
In Fig.~\ref{fig: t_scph}, we show the phonon bands and $\alpha^2 F(\omega)$ for three different $T_{\rm SCPH}$ values. 
From Fig.~\ref{fig: t_scph}(a), we notice that the phonon bands in the frequency region above 100 cm$^{-1}$ are not affected by the $T_{\rm SCPH}$ value. However, for the flat phonon band near 75 cm$^{-1}$ and the dispersive phonon bands below, the frequencies of the phonon bands, especially near the M point do increase as $T_{\rm SCPH}$ rises.
This results in the visible upward shift of the bottom of the Eliashberg spectral function as $T_{\rm SCPH}$ increases [see Fig.~\ref{fig: t_scph}(b)].

Finally, we note from Eqs.~(\ref{eq:a2F}), (\ref{eq:lambda}) and (\ref{eq:logW}) that $\lambda\propto\omega^{-1}$ and $\omega_{\rm log}\propto \log(\omega)$. 
Since $T_{\rm SCPH}$ only affects the phonon frequency, we can see why $\lambda$ drops as $\omega_{\rm log}$ rises due to the increase of $T_{\rm SCPH}$. 
Furthermore, due to the competing scaling relations of $\lambda$ and $\omega_{\rm log}$ with respect to $\omega$, $T_c$ of Sc$_6$FeTe$_2$ evaluated using Eq.~(\ref{eq:MAD}) does not significantly depend on $T_{\rm SCPH}$. 

\section*{ACKNOWLEDGMENTS}
M.-C. J. and G.-Y. G. acknowledge the support from the Ministry of Science and Technology and the National Center for Theoretical Sciences (NCTS) of the R.O.C. M.-C. J. was supported by the IPA Program, RIKEN. 
R. M. was supported by Grant-in-Aid for JSPS Fellows (No. 22KJ1028). 
R. A. was supported by JSPS KAKENHI Grant Number JP24H00190.

\bibliographystyle{apsrev}
\bibliography{SFT_SC_abbr}

\end{document}